\documentclass[12pt]{article}
\usepackage{fullpage}
\usepackage{cite}
\usepackage{amstext,amsfonts,amsbsy,eucal,amssymb}
\usepackage{epsfig}
\usepackage{psfrag}

\newcounter{definition}
\newenvironment{definition}{
  \stepcounter{definition}

  {\it definition} \thedefinition.
\begin{quotation}
}{
\end{quotation}
}

\def\thefigure{\arabic{section}.\arabic{figure}}
\def\theequation{\thesection.\arabic{equation}}
\def\appendix{
  \setcounter{section}{0}
  \setcounter{subsection}{0}
  \par
  \def\thesection{Appendix \Alph{section}}
  \def\theequation{\Alph{section}.\arabic{equation}}
  \def\thefigure{\Alph{section}.\arabic{figure}}
}
\makeatletter
\def\fnum@figure{Fig. \thefigure}
\@addtoreset{equation}{section}
\@addtoreset{figure}{section}
\makeatother

\date{
May 27,2002
}
\title{
Direct Calculation of Thermodynamic Quantities\\
for Heisenberg Model
} 
\author{
  Go {\sc Kato}\thanks{kato@monet.phys.s.u-tokyo.ac.jp}\, and  
  Miki {\sc Wadati}\thanks{wadati@phys.s.u-tokyo.ac.jp} \\
  Department of Physics, Graduate School of Science,\\
  University of Tokyo,\\
  Hongo 7-3-1, Bunkyo-ku, Tokyo 113-0033, Japan.
}

\begin{document}
\maketitle \setlength{\baselineskip}{1.8em}
\begin{abstract}
  The XXX Heisenberg model is studied at finite temperature.  The free
  energy is derived without recourse to Thermal Bethe Ansatz method and
  Quantum Transfer Matrix method. The result perfectly agrees with the
  free energy derived by Thermal Bethe Ansatz method. An explicit
  expression of the cluster expansion coefficient in arbitrary order
  is presented for the first time.
\end{abstract}

\newpage
\section{Introduction}
In a pioneer work~\cite{bethe}, H. Bethe solved the XXX Heisenberg
model.  The Hamiltonian of the model is written as
\begin{eqnarray}
  H\left(J,h\right) &=& -
  J\sum_{j=1}^{N}\left(S^x_{j}S^x_{j+1}+S^y_{j}S^y_{j+1}+S^z_{j}S^z_{j+1}\right) +
  2h\sum_{j=1}^{N}S_j^x +{\rm constatn},
\label{eq:Hamiltonian}
\end{eqnarray}
where $J$ is the coupling constant and $h$ is the external field.  The
energy of the system is given by
\begin{eqnarray}
  E_M &=& 2hM+\sum_{j=1}^{M}\frac{2J}{x_j^2+1},
\label{eq:energy_orijinal}
\end{eqnarray}
where $M$ is the number of up-spins and the variables $x_j$ are
required to satisfy the Bethe Ansatz (BA) equation,
\begin{eqnarray}
  \left(\frac{x_j+i}{x_j-i}\right)^N &=& \prod_{j'\neq j}\frac{x_{j }-
    x_{j'}+2i}{x_{j }-x_{j'}-2i}.
\label{eq:boundary_condition_orijinal}
\end{eqnarray}
It was shown \cite{bethe} that every eigenstate of this model
corresponds to a solution of this simultaneous equation.

The XXX Heisenberg model is the quantum integrable system.  Thermal
Bethe Ansatz method (TBA)~\cite{yang} and Quantum Transfer Matrix
method
(QTM)~\cite{koma,inoue_suzuki,wadati_akutsu,suzuki_akutsu_wadati,suzuki_nagao_wadati}
are the well-known methods to derive thermodynamic quantities for
quantum integrable systems.  TBA was originated by Yang and
Yang~\cite{yang}, who derived the thermodynamics of repulsive
$\delta$-function bose gas system.  Then, on the assumption of the
so-called string hypothesis, M.Takahashi~\cite{takahashi_a} applied
TBA to the Heisenberg model.  Recently, some analytical
results of thermodynamic quantities using QTM were presented by
A. Kl\"umper~\cite{kloemper_a,kloemper_b}.  Many pieces of concept
with respect to QTM had been introduced before then. The equivalence
of the results using TBA and QTM was proved by several
researchers~\cite{takahashi_b,takahashi_c,kloemper_a,kuniba}.

Our purpose is to derive the free energy of this system independent of
TBA and QTM.  In this paper, we use a method which we call a direct
method.  We start from the string hypothesis.  There is no other
mathematical assumptions except the string hypothesis.  In other
words, using the direct method, a problem based on the string
hypothesis is solved.  On the other hand, TBA is a method to solve the
problem using a definition of non-equilibrium state entropy. As a main
result of this paper, the direct method justifies TBA.

The direct method was used for the $\delta$-function bose system
\cite{go_b,go_d,go_e,go_f}. We apply the method to the Heisenberg model here.
We define $z$ and $\beta$ as $z\equiv e^{-2\beta h}$
and $\beta\equiv\frac1{k_BT}$.
The direct method enables us to express the cluster expansion
coefficients $b_n$ in arbitrary order explicitly,
\begin{eqnarray}
  \log\left[z^{N/2}{\rm Tr}\;e^{-\beta H\left(J,h\right)}\right]
  &\equiv& \sum_{n=1}^\infty b_nz^n,
\end{eqnarray}
and, the free energy,
\begin{eqnarray}
  \frac1N\log \left[ {\rm Tr}\;e^{-\beta H\left(J,h\right)} \right],
\end{eqnarray}
by use of a function which is a solution of a nonlinear integral
equation.  Thus, we can show that the expression of the free energy
perfectly agrees with the results of TBA~\cite{takahashi_b}.

The outline of this paper is the following. In \S2, we derive an
explicit expression of cluster expansion coefficients in all
orders. In \S3, we prove that the free energy can be expressed in
terms of a solution of a nonlinear integral equation. The last section
is devoted to the concluding remarks. Technical details of
calculations are summarized in Appendices A-H.

\section{The cluster expansion}
The string hypothesis is formulated as follows.  In the thermodynamic
limit, the energy of the Heisenberg model (\ref{eq:Hamiltonian}) can
be written as
\begin{eqnarray}
  E\left(J,h\right)
&=&
  2hM+\sum_{n,\alpha}\frac{2Jn}{\left(x^n_\alpha\right)^2+n^2}.
\label{eq:energy_string}
\end{eqnarray}
Here, the variables $x^n_\alpha$ are determined by
\begin{eqnarray}
  \left(\frac{x^n_\alpha+ni}{x^n_\alpha-ni}\right)^N &=&
  \prod_{(m,\beta)\neq
    (n,\alpha)}E_{nm}\left(x^n_\alpha-x^m_\beta\right),
\label{eq:boundary_condition_string}
\end{eqnarray}
where
\begin{eqnarray}
  E_{nm}\left(x\right) &=&
  \frac{\left(x-\left(n+m\right)i\right)\left(x-\left|n-m\right|i\right)}
  {\left(x+\left(n+m\right)i\right)\left(x+\left|n-m\right|i\right)}
  \prod_{l=0}^{\min(n,m)}
  \left(\frac{x+\left(n+m-2l\right)i}{x-\left(n+m-2l\right)i}\right)^2.
\end{eqnarray}
Equation (\ref{eq:boundary_condition_string}) is referred to as string ceter equation. Every eigenstate corresponds to a solution of these simultaneous
equations.  Taking a logarithm of both sides of
(\ref{eq:boundary_condition_string}), we obtain
\begin{eqnarray}
  2N\tan^{-1}\left(\frac {x^n_\alpha} n\right)
&=&
  2\pi I^n_\alpha
  +\sum_{(m,\beta)\neq
    (n,\alpha)}\Delta_{n,m}\left(x^n_\alpha-x^m_\beta\right),
\label{eq:mBA_equation}
\end{eqnarray}
where
\begin{eqnarray}
  \Delta_{n,m}\left(x\right)
\;=\;
  -2\tan\!\!\!{}^{-1}\!\!\left(\frac x {n+m}\right)
  -2\left(1+\delta_{nm}\right)
  \tan\!\!\!{}^{-1}\!\!\left(\frac x{\left|n-m\right|}\right)
  \!\!+\!\!\!\!\!\!\sum_{l=0}^{\min(n,m)}\!\!\!\!\!
  4\tan\!\!\!{}^{-1}\!\!\left(\frac x {n+m-2l} \right)\!\!.
\label{eq:def_Delta}
\end{eqnarray}
From now on, we call (\ref{eq:mBA_equation}) modified Bethe Ansatz
(mBA) equation.

The string hypothesis indicates that the partition function
is a sum of the Boltzmann weights with respect to integers(or half
integers) $\{I^n_\alpha\}$ which define the energy of the system,
\begin{eqnarray}
  z^{N/2}{\rm Tr}\;e^{-\beta H\left(J,h\right)} &=&
  \sum_{\{I^n_\alpha\}}e^{-\beta E\left(J,h\right)-2 \beta  h M}.
  \label{eq:partition_function_formal_sum}
\end{eqnarray}
When we take the thermodynamic limit, we can replace the summations
over discrete variables by integral,
\begin{eqnarray}
  \int e^{-\beta E\left(J,h\right) -2 \beta  h M} \prod dI^n_\alpha.
  \label{eq:partition_function_formal_int}
\end{eqnarray}
There are some points which we have to consider carefully.  The Fermi
statistics must be taken into account in this integral.  In this
case, the exclusion property leads to the conditions that $x^n_\alpha\neq
x^n_\beta$ and that a state specified by $\{x_\alpha^n\}$ and a state
specified by $\{x_\alpha^{\prime n}\}$ which satisfies
$x_\alpha^{\prime n}=x_{p\left(\alpha\right)}^n$ are the same, where
$p$ is a permutation.

To express a concrete form of
(\ref{eq:partition_function_formal_int}), we prepare 3 symbols.
\begin{definition}
  $N_A$ denotes the number of elements in a set $A$.
\end{definition}
\begin{definition}
  $\Theta\left(A\right)$ denotes all the patterns of division of a set
  $A$.  A pattern of division is represented as a set having elements
  each of which is a cluster.  The cluster is one of pieces into which
  a set $A$ is divided, and the cluster is regarded as a set.  For
  example, for $A=\{a,b,c\}$, $\Theta(A)$ means
\begin{eqnarray}
\Theta\left(\{a,b,c\}\right)
&=&
    \{\{a\},\{b\},\{c\}\},
    \{\{a\},\{b  ,  c\}\},\nonumber\\&&{}
    \{\{b\},\{c  ,  a\}\},
    \{\{c\},\{a  ,  b\}\},
    \{\{a  ,  b  ,  c\}\}.
\end{eqnarray}
On condition that the domain of $\Theta(n)$ is the natural number,
$n\in\mathbb{N}$, $\Theta(n)$ is interpreted as
$\Theta(\{1,2,\cdots,n\})$.
\end{definition}
\begin{definition}
  $\bar\Theta\left(B\right)$ denotes all the patterns of division of a
  set $B$, when each element of $B$ is a set and each pattern of
  division satisfies the following condition; all the sets which are
  included in each cluster have the same number of elements.  For
  example, we can show for $\sigma_1=\{a,b\}$, $\sigma_2=\{c \}$,
  $\sigma_3=\{d \}$,
\begin{eqnarray}
  \bar\Theta\left(\{\sigma_1,\sigma_2,\sigma_3\}\right)
&=&
    \{\{\sigma_1\},\{\sigma_2\},\{\sigma_3\}\},
    \{\{\sigma_1\},\{\sigma_2  ,  \sigma_3\}\}.
\end{eqnarray}
\end{definition}
Using the above symbols, the integral
(\ref{eq:partition_function_formal_int}) can be written explicitly in
the following form,
\begin{eqnarray}
  z^{N/2}{\rm Tr}\;e^{-\beta H\left(J,h\right)}
&=&
  1 +\sum_{n=1}^\infty \frac {z^n}{n!} \sum_{\theta \in\Theta(n)}
  \left[\prod_{\sigma\in\theta}N_\sigma ! \right]
  \sum_{\zeta\in\bar\Theta(\theta)} \left[
    \prod_{\theta'\in\zeta}(-1)^{N_{\theta'}-1}\left(N_{\theta'}-1\right)!
  \right]
  \nonumber\\&&{}
  \int 
  \left|\frac{\partial I}{\partial x}\right|_{N,\zeta}
  e^{-\beta E\left(\zeta\right)} \prod_{\theta'\in \zeta}d
  x_{\theta'}.
\label{eq:partition_function_real_int}
\end{eqnarray}
We explain functions, $\left|\frac{\partial I}{\partial
    x}\right|_{N,\zeta}$ and $E\left(\zeta\right)$ used in this
equation, where $\zeta\in\bar\Theta(\theta)$ and $\theta\in\Theta(n)$.
The Jacobian
$\left|\frac{\partial I}{\partial x}\right|_{N,\zeta}$ is constructed
as follows. First, we consider $\theta$ uniquely determined by the
condition $\zeta\in\bar\Theta(\theta)$. Second, we make correspondence
between each element $\sigma$ of $\theta$ and two suffixes
$(n,\alpha)$ for the variables $I^n_\alpha$ and $x^n_\alpha$
introduced in the string hypothesis. In this case, the condition
$N_\sigma=n$ is required.  Third, we consider the mBA equation for the
variables $\{I^n_\alpha\}$, $\{x^n_\alpha\}$ which are constructed out
of $\theta$.  Fourth, we introduce $x_{\theta'}$, $I_{\theta'}$, where
$\theta'$ is an element of $\zeta$.  Here, we have constraining
conditions; $x_{\theta'}=x^n_\alpha$ and $I_{\theta'}=I^n_\alpha$,
where $(n,\alpha)$ corresponds to $\sigma\in\theta'$. Finally,
$\left|\frac{\partial I}{\partial x}\right|_{N,\zeta}$ is the Jacobian
between $\{x_{\theta'}\}$ and $\{I_{\theta'}\}$ which satisfy
respectively the mBA equation made from $\theta$ and the constraining
conditions made from $\zeta$.  Explicit expression of such Jacobian
matrix is given in eq.(\ref{eq:def_Jacobian_start}). We show the above
procedure by an example. For the case,
\begin{eqnarray}
  \begin{array}{rclrclrcl}
  \zeta&=&\{\theta_1,\theta_2\},&
  \theta_1&=&\{\{1,2\}\},&
  \theta_2&=&\{\{3\},\{4\}\},
  \end{array}
\label{eq:example_condition}
\end{eqnarray}
we have
\begin{eqnarray}
  \theta
&=&
  \{\{1,2\},\{3\},\{4\}\},
\quad\quad\zeta
\;\in\;
\bar\Theta\left(\theta\right)
\nonumber\\
  2N\tan^{-1}\left(\frac{x_{\theta_1}}{2}\right)
&=&
  2\pi I_{\theta_1}+2\Delta_{2,1}\left(x_{\theta_1}-x_{\theta_2}\right)
\nonumber\\
  2N\tan^{-1}\left(x_{\theta_2}\right)
&=&
  2\pi I_{\theta_2}+\Delta_{1,2}\left(x_{\theta_2}-x_{\theta_1}\right)
\nonumber\\
  \left(2\pi\right)^2\left|\frac{\partial I}{\partial x}\right|_{N,\zeta}
&=&
  \frac{4N}{x_{\theta_1}^2+4}\frac{2N}{x_{\theta_2}^2+1}
  \nonumber\\&&{}
  -\frac{4N}{x_{\theta_1}^2+4}
  \left(
    \frac{2}{\left(x_{\theta_1}-x_{\theta_2}\right)^2+1}
    +\frac{6}{\left(x_{\theta_1}-x_{\theta_2}\right)^2+9}
  \right)
\nonumber\\&&{}
  -2\frac{2N}{x_{\theta_2}^2+1}
  \left(
    \frac{2}{\left(x_{\theta_2}-x_{\theta_1}\right)^2+1}
    +\frac{6}{\left(x_{\theta_2}-x_{\theta_1}\right)^2+9}
  \right).
\end{eqnarray}
The energy $E\left(\zeta\right)$ is expressed in terms of
$\{x^n_\alpha\}$ made from $\theta$,
\begin{eqnarray}
  E\left(\zeta\right)
&=&
  \sum_{n,\alpha}\frac{2Jn}{\left(x^n_\alpha\right)^2+n^2}.
\end{eqnarray}
For the case (\ref{eq:example_condition}), $E\left(\zeta\right)$ is given by
\begin{eqnarray}
  E\left(\zeta\right)
&=&
  \frac{2J\times2}{x_{\theta_1}^2+4}
  +2\frac{2J }{x_{\theta_2}^2+1}.
\end{eqnarray}
Note that the constraining conditions made from $\zeta$ have been used.

Now, we show how (\ref{eq:partition_function_real_int}) is derived.
We notice two important facts.
One is that, we must exclude unphysical states from the sum, when the
sum is represented as the integral in
(\ref{eq:partition_function_formal_int}). The unphysical state means
the state corresponding to $\{x^n_\alpha\}$ where some two elements of
$\{x^n_\alpha\}$ coincides, $x^n_\alpha= x^n_\beta$. Those states
should be excluded because the norm of the state constructed out of
$\{x^n_\alpha\}$ satisfying such a condition is zero.  In
eq.(\ref{eq:partition_function_real_int}),
\begin{eqnarray}
    \prod_{\theta'\in\zeta}(-1)^{N_{\theta'}-1}\left(N_{\theta'}-1\right)!
\label{eq:coodinate_state}
\end{eqnarray}
are factors which exclude the unphysical states.  In the expression
(\ref{eq:partition_function_real_int}), we add unphysical states in
some terms of series and subtract them in some other terms. Then, the
sum of positive terms and negative terms is zero with respect to
unphysical states. This mathematical manipulation enables us to write
the free energy explicitly like
(\ref{eq:partition_function_real_int}).  Let us explain this
manipulation for the case of $\zeta$ in (\ref{eq:example_condition})
in detail.  The states corresponding to $\zeta$ in
(\ref{eq:example_condition}) are unphysical states.  We define
$\zeta'=\{\{1,2\}\}, \{\{3\}\}, \{\{4\}\}$.  $\zeta'$ corresponds to
the results of a string center equation
(\ref{eq:boundary_condition_string}) which defines one two-string
$x_{1,2}$ and two one-strings $x_{3}$ $x_{4}$.  On the other hand,
$\zeta=\{\{1,2\}\},\{\{3\},\{4\}\}$ in (\ref{eq:example_condition})
corresponds to the results of the string center equation which satisfy
$x_{3}=x_{4}$. Therefore, the results corresponding to $\zeta$ are
part of the results corresponding to $\zeta'$. Here, in the sum with
respect to elements in $\bar\Theta$
(\ref{eq:partition_function_real_int}), both $\zeta$ and $\zeta'$ are
summed up. And, the sum of coordinates (\ref{eq:coodinate_state}) of
terms corresponding to both $\zeta$ and $\zeta'$ is zero. Therefore,
unphysical states corresponding to $\zeta$ do not affect the sum in
(\ref{eq:partition_function_real_int}).
The other important fact is that, we must divide the sum by the number of the
symmetry. Here, the symmetry means that when $\{x^n_\alpha\}$ is a
solution of the mBA equation, $\{x^{\prime n}_\alpha\}$ which
satisfies the relation $x_\alpha^{\prime
  n}=x_{p\left(\alpha\right)}^n$ is also a solution of the mBA
equation.  In this case, it is necessary to divide the sum by the
number of the symmetry because $\{x^n_\alpha\}$ and $\{x^{\prime
  n}_\alpha\}$ correspond to the same state.  In
eq.(\ref{eq:partition_function_real_int}),
\begin{eqnarray}
   \left[\prod_{\sigma\in\theta}N_\sigma ! \right]/n!
\end{eqnarray}
is a factor for the correction of the sum which takes such symmetry
into account.

We have met similar correction factors in the analysis of the
$\delta$-function bose gas system. The reason for the appearance of
the correction factors is common.  A detailed proof for the
$\delta$-function bose gas system case is given in \S3.1 \cite{go_d}.

In order to express the Jacobian
$\left|\frac{\partial I}{\partial x}\right|_{N,\zeta}$ explicitly, we
further define three symbols.
\begin{definition}
  $M_{B}$ denotes a number of elements in one of the sets. Here, all
  the elements of a set $B$ are sets, and all sets in $B$ have the
  same number of elements.  Then, $M_B$ is defined as the number of
  elements in one of the sets,
  \begin{eqnarray}
    M_\theta
  &=&
    N_\sigma,
  \end{eqnarray}
  where $\sigma\in\theta\in\zeta\in\bar\Theta\left(\theta'\right)$ and
  $\theta'\in\Theta\left(n\right)$.
\end{definition}
\begin{definition}
  $\Lambda\left(A\right)$ denotes all the patterns of connection of a
  set $A$.  Here, what we call a pattern of connection satisfies the
  following two conditions.  1) Any two elements of $A$ are
  connected or not. In other words, there is no multiple connection.
  2) There is no closed path in the connections.  Then, a pattern of
  connection is represented as a set of elements each of which
  corresponds to a connection. Here, an element corresponding to a
  connection is a set which consists of two elements connected by the
  connection.  For example, we have
  \begin{eqnarray}
    \Lambda\left(\left\{a,b,c\right\}\right)
  &=&
    \emptyset,
    \left\{\left\{b,c\right\}\right\},
    \left\{\left\{c,a\right\}\right\},
    \left\{\left\{a,b\right\}\right\},\nonumber\\&&{}
    \left\{\left\{a,b\right\},\left\{a,c\right\}\right\},
    \left\{\left\{b,a\right\},\left\{b,c\right\}\right\},
    \left\{\left\{c,a\right\},\left\{c,b\right\}\right\}.
  \end{eqnarray}
\end{definition}
\begin{definition}
  $G([\lambda,A])$ is an element of $\Theta(A)$, where
  $\lambda\in\Lambda(A)$. In other words, $G([\lambda,A])$ is a
  pattern of division of $A$. The pattern of division $\theta$
  satisfies the following conditions; 1) any two connected elements
  are in the same set in $\theta$,and 2) elements of $\theta$ are
  larger in number than elements of any other $\theta'\in\Theta(A)$
  satisfying 1).  For example, we can show that 
  \begin{eqnarray}
    G\left(
      \left[\left\{\left\{a,b\right\}\right\},\left\{a,b,c\right\}\right]
    \right)
  &=&
    \left\{\left\{a,b\right\},\left\{c\right\}\right\}.
  \end{eqnarray}
\end{definition}

By use of the above  three symbols, the Jacobian can be expressed as
\begin{eqnarray}
      \left(2\pi\right)^{N_{\zeta}}
  \left|\frac{\partial I}{\partial x}\right|_{N,\zeta}
  &=&
  \left[\prod_{\theta\in\zeta}N_\theta\right]^{-1}
  \sum_{\lambda\in\Lambda\left(\zeta\right)}
  \left[\prod_{\left\{\theta,\theta'\right\}\in\lambda}
    -N_{\theta}N_{\theta'}
    K_{M_{\theta},M_{\theta'}}\left(x_{\theta}-x_{\theta'}\right)
  \right]
  \nonumber\\&&{}\times
     \prod_{\zeta'\in G\left(\left[\lambda,\zeta\right]\right)}
   \left[
     \sum_{\theta\in\zeta'}
     \frac{2N_\theta M_\theta N}{M_\theta^2+x^2_\theta}
   \right],
\label{eq:def_jacobian_n}
\end{eqnarray}
where $ K_{n,m}\left(x\right)\equiv d\Delta_{n,m}(x)/dx$.  A proof
of eq.(\ref{eq:def_jacobian_n}) is given in \ref{sec:proof_det}.

At last, we show the explicit expression of the cluster expansion.
Taking a logarithm of both sides of
(\ref{eq:partition_function_real_int}) and rewriting the r.h.s., we
get
\begin{eqnarray}
  \log\left[z^{N/2}{\rm Tr}\;e^{-\beta H\left(J,h\right)}\right] &=&
  \sum_{n=1}^\infty \frac {z^n}{n!} \sum_{\theta \in\Theta(n)}
  \left[\prod_{\sigma\in\theta}N_\sigma !\right]
  \sum_{\zeta\in\bar\Theta(\theta)} \left[
    \prod_{\theta'\in\zeta}(-1)^{N_{\theta'}-1}\left(N_{\theta'}-1\right)!
  \right] \nonumber\\&&{}
  \int
  \left|\frac{\partial I}{\partial x}\right|_{cN,\zeta}
  e^{-\beta E\left(\zeta\right)} \prod_{\theta'\in \zeta}d x_{\theta'}.
\label{eq:log_parititon_function}
\end{eqnarray}
A proof is given in \ref{sec:proof_log}.  Here,
$\left|\frac{\partial I}{\partial x}\right|_{cN,\zeta}$ is the first
order term of
$\left|\frac{\partial I}{\partial x}\right|_{N,\zeta}$ regarded as a
polynomial with respect to the number of spins, $N$. From the
expression (\ref{eq:log_parititon_function}), the cluster expansion
coefficient $b_n$ is given by
\begin{eqnarray}
b_n&=&
  \frac {1}{n!}
  \sum_{\theta \in\Theta(n)}
  \left[\prod_{\sigma\in\theta}N_\sigma !\right]
  \sum_{\zeta\in\bar\Theta(\theta)} \left[
    \prod_{\theta'\in\zeta}(-1)^{N_{\theta'}-1}\left(N_{\theta'}-1\right)!
  \right] \nonumber\\&&{}
  \int
  \left|\frac{\partial I}{\partial x}\right|_{cN,\zeta}
  e^{-\beta E\left(\zeta\right)} \prod_{\theta'\in \zeta}d x_{\theta'},
\end{eqnarray}
where
 the explicit form of
$\left|\frac{\partial I}{\partial x}\right|_{cN,\zeta}$
is
\begin{eqnarray}
    \left({2\pi}
    \right)^{N_{\zeta}}
    \!\!
  \left|\frac{\partial I}{\partial x}\right|_{cN,\zeta}
\;=\;
  \left[
    \prod_{\theta\in\zeta}N_\theta
  \right]^{-1}
  \!\!\!\!\sum_{\lambda\in\Lambda_c\left(\zeta\right)}
  \left[\prod_{\left\{\theta,\theta'\right\}\in\lambda}
    \!\!-N_{\theta}N_{\theta'}
    K_{M_{\theta},M_{\theta'}}\left(x_{\theta}-x_{\theta'}\right)
  \right]
     \sum_{\theta\in\zeta}
     \frac{2N_\theta M_\theta N}{M_\theta^2+x^2_\theta}.
\label{eq:def_jacobian_nc}
\end{eqnarray}
The symbol $\Lambda_c\left(A\right)$ is defined as follows.
\begin{definition}
  $\Lambda_c\left(A\right)$ is a subset of $\Lambda\left(A\right)$,
  where any element $\lambda\in\Lambda_c\left(A\right)$ satisfies the
  condition $N_{G[\lambda,A]}=1$. For example, we can show that
  \begin{eqnarray}
    \Lambda_c\left(\left\{a,b,c\right\}\right)
  &=&
    \left\{\left\{a,b\right\},\left\{a,c\right\}\right\},
    \left\{\left\{b,a\right\},\left\{b,c\right\}\right\},
    \left\{\left\{c,a\right\},\left\{c,b\right\}\right\}.
  \end{eqnarray}
\end{definition}
A derivation of the expression
$\left|\frac{\partial I}{\partial x}\right|_{cN,\zeta}$
in (\ref{eq:def_jacobian_nc}) is given in \ref{sec:proof_struct_jacobian}.

\section{The free energy}
From eqs.(\ref{eq:partition_function_real_int}) and
(\ref{eq:log_parititon_function}), we can easily show that
\begin{eqnarray}
  z^{1/2}\left[{\rm Tr}\;e^{-\beta H\left(J,h\right)}\right]^{1/N}
  &=& 1+\sum_{n=1}^\infty \frac {z^n}{n!} \sum_{\theta \in\Theta(n)}
  \left[\prod_{\sigma\in\theta}N_\sigma !\right]
  \sum_{\zeta\in\bar\Theta(\theta)} \left[
    \prod_{\theta'\in\zeta}(-1)^{N_{\theta'}-1}\left(N_{\theta'}-1\right)!
  \right] \nonumber\\&&{}
  \int
  \left|\frac{\partial I}{\partial x}\right|_{1,\zeta}
  e^{-\beta E\left(\zeta\right)} \prod_{\theta'\in \zeta}d
  x_{\theta'},
\end{eqnarray}
where
$\left|\frac{\partial I}{\partial x}\right|_{1,\zeta}$ means
$\left|\frac{\partial I}{\partial x}\right|_{N,\zeta}$ in which $N$ is
replaced with $1$.

Now, we introduce a function $u\left(x\right)$,
\begin{eqnarray}
  u\left(x\right)
&\equiv&
  1+\sum_{n=1}^\infty  {z^n}u_n\left(x\right),
\label{eq:definition_u}
\end{eqnarray}
where $u_n(x)$ is given by

\begin{eqnarray}
  u_n\left(x\right)
&\equiv&
  \frac 1{n!} \sum_{\theta \in\Theta(n)}
  \left[\prod_{\sigma\in\theta}N_\sigma !\right]
  \sum_{\zeta\in\bar\Theta(\theta)} \left[
    \prod_{\theta'\in\zeta}(-1)^{N_{\theta'}-1}\left(N_{\theta'}-1\right)!
  \right]
 \nonumber\\&&{}
  \int
  \left|\frac{\partial I}{\partial x}\right|_{1,\zeta}\left(x\right)
  e^{-\beta E\left(\zeta\right)} \prod_{\theta'\in \zeta}d
  x_{\theta'}
\label{eq:definition_un}
\end{eqnarray}
with the modified Jacobian
$\left|\frac{\partial I}{\partial x}\right|_{1,\zeta}(y)$ defined
as
\begin{eqnarray}
      \left(2\pi\right)^{N_{\zeta}}
    \left|\frac{\partial I}{\partial x}\right|_{1,\zeta}\left(y\right)
&\equiv& 
  \left[\prod_{\theta\in\zeta}N_\theta\right]^{-1}
  \sum_{\lambda\in\Lambda\left(\zeta\right)}
  \left[\prod_{\left\{\theta,\theta'\right\}\in\lambda}
    -N_{\theta}N_{\theta'}
    K_{M_{\theta},M_{\theta'}}\left(x_{\theta}-x_{\theta'}\right)
  \right]
  \nonumber\\&&{}
     \prod_{\zeta'\in G\left(\left[\lambda,\zeta\right]\right)}
   \left[
     \sum_{\theta\in\zeta'}
     \frac{2N_\theta M_\theta}{M_\theta^2+\left(x_\theta-y\right)^2}
   \right].
\label{eq:def_modified_jacobian}
\end{eqnarray}
Comparing this definition with (\ref{eq:def_jacobian_n}), it is
readily seen that
$\left|\frac{\partial I}{\partial x}\right|_{1,\zeta} =
\left|\frac{\partial I}{\partial x}\right|_{1,\zeta}\left(0\right)$.
Then, it is clear that we can get a relation
\begin{eqnarray}
  \left[{\rm Tr}\;e^{-\beta H\left(J,h\right)}\right]^{1/N}
&=&
  u\left(0\right)z^{-1/2}.
  \label{eq:relation_u_jacobian}
\end{eqnarray}

It can be shown (see \ref{sec:proof_u_n} for a proof) that the function $u_n(x)$ satisfies the following
recursion relation,
\begin{eqnarray}
  u_{n+1}\left(x\right) &=& 
    \oint_{0+}\left[\frac{\exp\left(-\frac{2J\beta}{(y+i)^2+1}\right)}{x-y-2i}
    +\frac{\exp\left(-\frac{2J\beta}{(y-i)^2+1}\right)}{x-y+2i}\right]
  \nonumber\\{}&&
   \sum_{\theta\in\Theta\left(n\right)}\frac{N_\theta!}{n!}
   \left[
     \prod_{\sigma\in\theta}-N_\sigma!u_{N_\sigma}\left(y\right)
   \right]
   \frac{dy}{2\pi i}
\label{eq:relation_u_n}
\\
  u_1\left(x\right) &=& 1
  +\oint_{0+}\left[\frac{\exp\left(-\frac{2J\beta}{(y+i)^2+1}\right)}{x-y-2i}
    +\frac{\exp\left(-\frac{2J\beta}{(y-i)^2+1}\right)}{x-y+2i}\right]
      \frac{dy}{2\pi i},
\label{eq:relation_u_1}
\end{eqnarray}
where $n$ is the natural number. 
  Using this recursion relation we obtain a
relation
\begin{eqnarray}
  u\left(x\right) &=& z+1
  +\oint_{0+}\left[\frac{\exp\left(-\frac{2J\beta}{(y+i)^2+1}\right)}{x-y-2i}
    +\frac{\exp\left(-\frac{2J\beta}{(y-i)^2+1}\right)}{x-y+2i}\right]
      \frac{z}{u\left(y\right)}\frac{dy}{2\pi i}.
\label{eq:recursive_relation_u_n}
\end{eqnarray}
These results, (\ref{eq:relation_u_jacobian}) and
(\ref{eq:recursive_relation_u_n}), are the same as those of
\cite{takahashi_b} in the XXX Heisenberg model limit.

\section{Conclusion}
We have shown a method, which we call direct method, to derive the free energy of the XXX Heisenberg
model using the BA equation. The cluster expansion coefficient in
arbitrary order is given for the first time.  The expression of the
free energy perfectly agrees with that by TBA.  It is remarkable that
the free energy is obtained without recourse to TBA and QTM. On the
other hand, there remains a problem. We have started from the string
hypothesis in this paper. Therefore, it is still a challenging problem
to derive each cluster expansion coefficient only from the BA
equation. This problem is under investigation.

We think, however, that these results have theoretical significances.
What we have done in this paper is that we calculate a certain
summation using combinatorial argument with mathematical
justification.  The summation may contain a problem which comes from
the string hypothesis.  On the contrary, TBA solve this problem as
follows.  We define a non-equilibrium entropy, and minimize the free
energy using such entropy. Then, the condition of minimization gives
the summation. In other words, TBA entirely relies on the physical
definition of the entropy.  In this paper, it is proved that the two
results by TBA and by the direct method are the same. Therefore, in
addition to the $\delta$-function bose gas, for the XXX Heisenberg
model, the direct method gives the mathematical justification of TBA.
To summarize, it is suggested that TBA can be systematically proved
for many other integrable systems using combinatorial argument
presented here.

\newpage

\appendix

\section{
  Explicit form of Jacobian
  $\left|\frac{\partial I}{\partial x}\right|_{N,\zeta}$ }
\label{sec:proof_det}
In this appendix, we prove (\ref{eq:def_jacobian_n}). Recall that
$\left|\frac{\partial I}{\partial x}\right|_{N,\zeta}$ is the Jacobian
defined for $\zeta\in\bar\Theta(\theta)$, where $\theta\in\Theta(n)$.
The Jacobian matrix can be rewritten as follows; the Jacobian matrix
has the order $N_\zeta\times N_\zeta$, and each element of the matrix
is
\begin{eqnarray}
  2\pi
  \frac{\partial I_{\theta'}}{\partial x_{\theta}}
  &=&
  \left\{
    \begin{array}{ll}
      \frac{2M_\theta N}{M_\theta^2+x^2_\theta}
      -\sum_{\theta''\neq\theta} N_{\theta''}
      K_{M_\theta ,M_{\theta''}}\left(x_\theta-x_{\theta''}\right)
      & \makebox{if $\theta=\theta'$}\\[4mm]
      N_{\theta}
      K_{M_\theta, M_{\theta' }}\left(x_\theta-x_{\theta' }\right)
      & \makebox{if $\theta\neq\theta'$},
    \end{array}
  \right.
  \label{eq:def_Jacobian_start}
\end{eqnarray}
where $\theta,\theta',\theta''\in\zeta$ and $ K_{n,m}\left(x\right)\equiv d\Delta_{n,m}(x)/dx$.

Here, we generalize the setting of the problem. We define $N\times N$
matrix
\begin{eqnarray}
  A_{n,m}
  &=&
  \left\{
    \begin{array}{ll}
      e_n
      -\sum_{k\neq n} c_k
      a_{n,k}
      & \makebox{if $n=m$}\\[4mm]
      c_n
      a_{n,m}
      & \makebox{if $n\neq m$},
    \end{array}
  \right.
  \label{eq:matrix_a}
\end{eqnarray}
where $\{a_{n,m}\}$, $\{c_n\}$ and $\{e_n\}$ are arbitrary series and
$a_{n,m}$ is symmetric, $a_{n,m}=a_{m,n}$.  This matrix is a
generalized form of the Jacobian matrix (\ref{eq:def_Jacobian_start}).
Therefore, what we have to prove becomes that the determinant of
$A_{n,m}$ is
\begin{eqnarray}
  \left|A_{n,m}\right|
  &=&
  \left[\prod_{n=1}^{N}c_n\right]^{-1}
  \sum_{\lambda\in\Lambda\left(\left\{1,\cdots,N\right\}\right)}
  \left[\prod_{\left\{n,m\right\}\in\lambda}
    -c_nc_m
    a_{n,m}
  \right]
  \prod_{\sigma\in G\left(\left[\lambda,\{1,\cdots,N\}\right]\right)}
  \left[
    \sum_{n\in\sigma}
    c_ne_n
  \right].
  \label{eq:determinant_a}
\end{eqnarray}

First, we prove that each term of the determinant $|A_{n,m}|$ can be
considered as a pattern of connection
$\lambda\in\Lambda\left(\left\{1,\cdots,N\right\}\right)$.  When we
regard the determinant as a multi-variable polynomial with respect to
$\{a_{n,m}\}$ and regard $a_{n,m}$ as a connection between $n$ and
$m$, we can consider each term of the polynomial as a pattern of
connection $\lambda$.  In other words, we can prove the following two
facts. First, there is no multiple connection. That is to say, there
is no $a_{n,m}$ to the power of 2 or more in any term of the
polynomial.  Second, there is no closed path made from
connections. That is to say, there is no product
$a_{p(1),p(2)}a_{p(2),p(3)}\cdots a_{p(M-1),p(M)}a_{p(M),p(1)}$ in any
term, where $p$ is a permutation.

To prove these two facts, we use
the method of false position.
  We assume there is a term that has at
least one closed path through $M$ connections.  Without loss of
generality, the above assumption can be rewritten that there is a
product $a_{1,2}a_{2,3}\cdots a_{M-1,M}a_{M,1}$ in any term.  Then, we
introduce a matrix $A'_{n,m}$ such that
\begin{eqnarray}
  A'_{n,m}
  &=&
  \left\{
    \begin{array}{ll}
      -\sum_{k\neq n} c_k
      a_{n,k}
      & \makebox{if $n=m\leq M$}\\[4mm]
      e_n-\sum_{k\neq n} c_k
      a_{n,k}
      & \makebox{if $n=m > M$}\\[4mm]
      0
      & \makebox{if $n\leq M < m$ or $m\leq M < n$}\\[4mm]
      c_n
      a_{n,m}
      & \makebox{other}.
    \end{array}
  \right.
  \label{eq:matrix_a'}
\end{eqnarray}
Now, we regard $|A'_{n,m}|$ as a multi-variable polynomial with
respect to $\{a_{n,m}\}$.  From the definition of determinant, it is
easily shown that the term with a product $a_{1,2}a_{2,3}\cdots
a_{M-1,M}a_{M,1}$ in $|A_{n,m}|$ has the same coefficient as a term
with such product in $|A'_{n,m}|$.  However, $|A'_{n,m}|$ is
identically $0$ because the matrix $A'_{n,m}$ is linearly
dependent. Therefore, there is no term containing a product
$a_{1,2}a_{2,3}\cdots a_{M-1,M}a_{M,1}$. Substituting 2 for $M$, we can
prove that there is no $a_{n,m}$ to the power of 2 or more in any term
of the polynomial. Then, we can consider each term of the matrix
$|A_{n,m}|$ as a pattern of connection $\lambda$.

Second, we find where a term considered as a pattern of connection is
formed. From the definition, we can write the determinant $|A_{n,m}|$
as
\begin{eqnarray}
  \left|A_{n,m}\right|
  &=&
  \sum_{p\in P_N}{\rm sgn} p\;\cdot\;\prod_{n=1}^{N}A_{n,p\left(n\right)},
\end{eqnarray}
where $P_N$ is a permutation group with respect to $\{1,\cdots,N\}$.
It can be shown that any term considered as a pattern of connection is
in an expansion of a product only in the case $p$ is the identity
permutation. The following is a proof of the fact by the method of false position.  If
there is such term in the case $p$ is not the identity permutation,
there exist $n_0$ and $l_0$ which satisfy $p^{l_0}(n_0)= n_0$,
$p^{l}(n_0)\neq n_0$ and $l_0\neq0$, where $l_0>l$ and $p^m(n)$ means
that $p$ operates on $n$ for $m$ times, $p(p(\cdots p(n)\cdots))$.
Therefore, we can rewrite $\prod_{n=1}^{N}A_{n,p\left(n\right)}$ as
\begin{eqnarray}
  \left[\prod_{l=0}^{l_0-1}c_{p^l(n_0)}a_{p^l(n_0),p^{l+1}(n_0)}\right]
\prod_{n\neq p^l(n_0)}A_{n,p\left(n\right)}.
\end{eqnarray}
This term clearly has a closed path, and conflict with the definition of the $p$.

Finally, we can rewrite $|A_{n,m}|$ as follows.  We expand
\begin{eqnarray}
  \prod_{n=1}^{N}A_{n,n},
\end{eqnarray}
and regard it as a multi-variable polynomial with respect to
$\{a_{n,m}\}$.  $\left|A_{n,m}\right|$ is a sum of terms which are in
this polynomial and can be considered as a pattern of connection. This
means that the term has no multiple connection and no closed path.
This sum means the r.h.s. of (\ref{eq:determinant_a}).

\section{
  A proof of (\ref{eq:log_parititon_function})
}
\label{sec:proof_log}
In this appendix, we prove (\ref{eq:log_parititon_function}) from
(\ref{eq:partition_function_real_int}).

We define $Z_n$ by
\begin{eqnarray}
  z^{N/2}{\rm Tr}\;e^{-\beta H\left(J,h\right)}
  &\equiv&
  1 +\sum_{n=1}^\infty Z_nz^n.
  \label{eq:tmp_a_06}
\end{eqnarray}
From (\ref{eq:partition_function_real_int}) we get
\begin{eqnarray}
  Z_n
  &=&
  \frac {1}{n!}
  \!\!\sum_{\theta \in\Theta(n)}\!
  \left[\prod_{\sigma\in\theta}N_\sigma ! \right]
  \!\!\sum_{\zeta\in\bar\Theta(\theta)}\!
  \left[
    \prod_{\theta'\in\zeta}(-1)^{N_{\theta'}-1}\!
    \left(N_{\theta'}-1\right)!
  \right]
  \int
  \left|\frac{\partial I}{\partial x}\right|_{N,\zeta}\!\!
  e^{-\beta E\left(\zeta\right)} \prod_{\theta'\in \zeta}d
  x_{\theta'},
\end{eqnarray}
where
$\left|\frac{\partial I}{\partial x}\right|_{N,\zeta}$ is the Jacobian
defined for $\zeta$.  It is convenient to introduce $f(\zeta)$ as
\begin{eqnarray}
  f\left(\zeta\right)
  &\equiv&
  \int
  \left|\frac{\partial I}{\partial x}\right|_{cN,\zeta}
  e^{-\beta E\left(\zeta\right)} \prod_{\theta'\in \zeta}d x_{\theta'}.
  \label{eq:tmp_a_05}
\end{eqnarray}
Using this function and the relation (\ref{eq:struct_jacobian}) proved
in \ref{sec:proof_struct_jacobian}, we can rewrite $Z_n$ as
\begin{eqnarray}
  Z_n&=&\frac {1}{n!} \sum_{\theta \in\Theta(n)}
  \left[\prod_{\sigma\in\theta}N_\sigma ! \right]
  \sum_{\zeta\in\bar\Theta(\theta)} \left[
    \prod_{\theta'\in\zeta}(-1)^{N_{\theta'}-1}\left(N_{\theta'}-1\right)!
  \right]
  \sum_{\xi\in\Theta\left(\zeta\right)}\prod_{\zeta'\in\xi}f\left(\zeta'\right).
\label{eq:tmp_a_01}
\end{eqnarray}

We also define $b_n$ by
\begin{eqnarray}
  \log\left[1 +\sum_{n=1}^\infty Z_nz^n\right]
  &\equiv&
  \sum_{n=1}^\infty b_nz^n.
  \label{eq:tmp_a_04}
\end{eqnarray}
We can easily show the following relation between $b_n$ and $Z_n$,
\begin{eqnarray}
  b_n
  &=&
  \frac1{n!}
  \sum_{\theta\in\Theta\left(n\right)}
  \left(N_\theta-1\right)!(-1)^{N_\theta-1}
  \prod_{\sigma\in\theta}N_\sigma!Z_{N_\sigma}.
  \label{eq:tmp_a_02}
\end{eqnarray}

Substituting (\ref{eq:tmp_a_01}) for $Z_n$ in (\ref{eq:tmp_a_02}),
 we get
\begin{eqnarray}
  b_n
  &=&
  \frac1{n!}
  \sum_{\theta\in\Theta\left(n\right)}
  \left(N_\theta-1\right)!(-1)^{N_\theta-1}
\nonumber\\&&{}
  \prod_{\sigma\in\theta}
  \left[
    \sum_{\theta' \in\Theta(\sigma)}
    \left[\prod_{\sigma'\in\theta'}N_{\sigma'} ! \right]
    \sum_{\zeta\in\bar\Theta(\theta')} \left[
      \prod_{\theta''\in\zeta}(-1)^{N_{\theta''}-1}\left(N_{\theta''}-1\right)!
    \right]
    \sum_{\xi\in\Theta\left(\zeta\right)}\prod_{\zeta'\in\xi}f\left(\zeta'\right)
  \right]
\nonumber\\&=&
  \frac1{n!}
  \sum_{\theta\in\Theta\left(n\right)}
  \left[\prod_{\sigma\in\theta}N_{\sigma} ! \right]
  \sum_{\zeta\in\bar\Theta(\theta)} \left[
    \prod_{\theta'\in\zeta}(-1)^{N_{\theta'}-1}\left(N_{\theta'}-1\right)!
  \right]
  \sum_{\xi\in\Theta\left(\zeta\right)}
  \left[\prod_{\zeta'\in\xi}f\left(\zeta'\right)\right]
\nonumber\\&&{}
  \sum_{\eta \in\Theta(\xi)}
  \left(N_\eta-1\right)!(-1)^{N_\eta-1}
\nonumber\\&=&
  \frac1{n!}
  \sum_{\theta\in\Theta\left(n\right)}
  \left[\prod_{\sigma\in\theta}N_{\sigma} ! \right]
  \sum_{\zeta\in\bar\Theta(\theta)} \left[
    \prod_{\theta'\in\zeta}(-1)^{N_{\theta'}-1}\left(N_{\theta'}-1\right)!
  \right]
  f\left(\zeta\right).
  \label{eq:tmp_a_03}
\end{eqnarray}
In the first equality, we have only performed the substitution.  In
the second equality, we have done the following manipulation.  We
regard each side of the equality as the sum with respect to elements
in a set which satisfies some conditions. Therefore, we may change the
order of the sums. Here, we show a simple example of ``change of
order'',
\begin{eqnarray}
  \sum_{n=1}^{N}\sum_{m=1}^{n}f\left(\left\{n,m\right\}\right)
  &=&
  \sum_{m=1}^{N}\sum_{n=m}^{N}f\left(\left\{n,m\right\}\right).
\end{eqnarray}
Similar to this example, both sides of the second equality are the sum
of the same function with respect to elements in the same set.  In the
third equality, we have simplified the last term of the l.h.s.  using
a relation,
\begin{eqnarray}
  \sum_{\theta \in\Theta(\sigma)}
  \left(N_\theta-1\right)!(-1)^{N_\theta-1}&=&\delta_{N_\sigma,1}.
  \label{eq:combinatorical_relation_1}
\end{eqnarray}

Finally, combining (\ref{eq:tmp_a_06}), (\ref{eq:tmp_a_05}),
(\ref{eq:tmp_a_04}) and (\ref{eq:tmp_a_03}) we arrive at the equation
(\ref{eq:log_parititon_function}).

\section{
The structure of the Jacobian 
$\left|\frac{\partial I}{\partial x}\right|_{N,\zeta}$
}
\label{sec:proof_struct_jacobian}
In this appendix, we prove
\begin{eqnarray}
  \left|\frac{\partial I}{\partial x}\right|_{N,\zeta}
&=&
  \sum_{\xi\in\Theta\left(\zeta\right)}\prod_{\zeta'\in\xi}
  \left|\frac{\partial I}{\partial x}\right|_{cN,\zeta'}
  \label{eq:struct_jacobian}
\end{eqnarray}
and eq.(\ref{eq:def_jacobian_nc}) from
eq.(\ref{eq:def_jacobian_n}).  We here repeat that
$\left|\frac{\partial I}{\partial x}\right|_{cN,\zeta}$ is the first
order term of
$\left|\frac{\partial I}{\partial x}\right|_{N,\zeta}$ with respect to
$N$.

Eq.(\ref{eq:def_jacobian_n}) can be rewritten as
\begin{eqnarray}
  \left({2\pi}\right)^{N_\zeta}
  \left|\frac{\partial I}{\partial x}\right|_{N,\zeta}
  &=&
  \sum_{\xi\in\Theta\left(\zeta\right)}
  \prod_{\zeta'\in\xi}
  \left[
    \left[\prod_{\theta\in\zeta'}N_\theta\right]^{-1}
    \sum_{\lambda\in\Lambda_c\left(\zeta'\right)}
    \left[\prod_{\left\{\theta,\theta'\right\}\in\lambda}
      -N_{\theta}N_{\theta'}
      K_{M_{\theta},M_{\theta'}}\left(x_{\theta}-x_{\theta'}\right)
    \right]\right.
\nonumber\\&&{}
  \left.
    \left.
      \sum_{\theta\in\zeta' N}
      \frac{2N_\theta M_\theta}{M_\theta^2+x^2_\theta}
    \right.
  \right].
  \label{eq:tmp_c_01}
\end{eqnarray}
In this equality, we have changed the order of the sums.  We can
easily show the relation (\ref{eq:def_jacobian_nc}) from this
expression.  Then, eq.(\ref{eq:tmp_c_01}) and
eq.(\ref{eq:def_jacobian_nc}) prove (\ref{eq:struct_jacobian}).

\section{
 A proof of (\ref{eq:relation_u_n}) and (\ref{eq:relation_u_1})
}
\label{sec:proof_u_n}
In this appendix, using the mathematical induction we prove the relations
with respect to $u_n(x)$, (\ref{eq:relation_u_n}) and
(\ref{eq:relation_u_1}).  The definition of $u_n(x)$ is given in
eq.(\ref{eq:definition_un}).

We can change the r.h.s of (\ref{eq:relation_u_1}) into 
\begin{eqnarray}
  \int\frac{2}{1+\left(x-y\right)^2}
  \exp\left(-\frac{2J\beta}{y^2+1}\right)
  \frac{dy}{2\pi i}.
\end{eqnarray}
This is the same as the definition of $u_1(x)$.  Thus, we
have shown the relation (\ref{eq:relation_u_1}).

 We suppose (\ref{eq:relation_u_n}) and (\ref{eq:relation_u_1}) hold for
 $1\leq n<n_0$, where $n_0$ is a natural number. For simplicity, $n_0$ is merely written as
$n$ in what follows.  Then, the function $u_m\left(x\right)$ has poles
only at $x=\pm2i$, and satisfies
$\lim_{\left|x\right|\rightarrow\infty}u_m(x)=\delta_{m,1}$, where $1\leq
m\leq n$. By use of these two properties, the r.h.s. of
(\ref{eq:relation_u_n}) becomes
\begin{eqnarray}
&&\nonumber
  \oint_{0+}\left[
    \frac{\exp\left(-\frac{2J\beta}{(y+i)^2+1}\right)}{x-y-2i}
    +\frac{\exp\left(-\frac{2J\beta}{(y-i)^2+1}\right)}{x-y+2i}
  \right]
  \sum_{\theta\in\Theta\left(n\right)}\frac{N_\theta!}{n!}
  \left[
    \prod_{\sigma\in\theta}-N_\sigma!u_{N_\sigma}\left(y\right)
  \right]
  \frac{dy}{2\pi i}
\\&=&
  \sum_{\theta\in\Theta\left(n\right)}\frac{N_\theta!}{n!}
  \int_{-\infty}^{\infty}
  \left[
    \frac{\prod_{\sigma\in\theta}-N_\sigma!u_{N_\sigma}\left(y-i\right)}
    {x-y-i}
    -\frac{\prod_{\sigma\in\theta}-N_\sigma!u_{N_\sigma}\left(y+i\right)}
    {x-y+i}
  \right]
  e^{-\frac{2J\beta}{y^2+1}}
  \frac{dy}{2\pi i}
\nonumber{}\\&&%
  -\sum_{\theta\in\Theta\left(n\right)}\frac{N_\theta!}{n!}
  \int_{C_+}
  \frac{\prod_{\sigma\in\theta}-N_\sigma!u_{N_\sigma}\left(y\right)}{x-y-2i}
  e^{-\frac{2J\beta}{(y+i)^2+1}}
  \frac{dy}{2\pi i}
\nonumber{}\\&&%
  +\sum_{\theta\in\Theta\left(n\right)}\frac{N_\theta!}{n!}
  \int_{C_-}
  \frac{\prod_{\sigma\in\theta}-N_\sigma!u_{N_\sigma}\left(y\right)}{x-y+2i}
  e^{-\frac{2J\beta}{(y-i)^2+1}}
  \frac{dy}{2\pi i}.
  \label{eq:tmp_f_01}
\end{eqnarray}
In this equality we have changed the path of integration.  Both $C_+$
and $C_-$ are the paths of integration which counterclockwise surround
the region $\Im y>0$. When these paths are in the neighborhood of real
axis, $C_+$ passes above the real axis and $C_-$ passes below the real
axis.  We separate (\ref{eq:tmp_f_01}) into two parts, and calculate
them separately.

First, we consider the first term of (\ref{eq:tmp_f_01}) which is
rewritten as
\begin{eqnarray}
  \sum_{\theta\in\Theta\left(n\right)}\frac{N_\theta!}{n!}
  \int_{-\infty}^{\infty}
  \sum_{s=\pm1}
  \frac{s\prod_{\sigma\in\theta}-N_\sigma!u_{N_\sigma}\left(y-si\right)}
  {s-\left(y-x\right)i}
  e^{-\frac{2J\beta}{y^2+1}}
  \frac{dy}{2\pi}.
  \label{eq:tmp_f_02}
\end{eqnarray}
The integral diverges in case that we do not sum up with respect to
$s$.  However, the sum turns out not to play an important role in the
following calculation.  Hence, to shorten the expression, a symbol
$\sum_{s=\pm1}$ is omitted for a while. Now, by use of the relation
(\ref{eq:simplification_prod_u_n}) proved in \ref{sec:proof_wrt_u_n},
eq.(\ref{eq:tmp_f_02}) is rewritten as
\begin{eqnarray}
&&
  \frac{1}{n!}
  \int_{-\infty+s\delta i}^{\infty+s\delta i}
  \frac{dy}{2\pi}
  \sum_{\theta\in\Theta\left(n\right)}
  \prod_{\sigma\in\theta}
  \left[
    -\sum_{\theta' \in\Theta(\sigma)}
    \left[\prod_{\sigma'\in\theta'}N_{\sigma'}!\right]
    \sum_{\zeta\in\bar\Theta(\theta')}
    \left[
      \prod_{\theta''\in\zeta}
      (-1)^{N_{\theta''}-1}\frac{\left(N_{\theta''}-1\right)!}{N_{\theta''}}
    \right]
  \right.
\nonumber\\&&{}
  \left.
    \left[
      \prod_{\theta''\in \zeta}
      \int_{-\infty}^{\infty}
      \frac{dx_{\theta''}}{2\pi}
    \right]
    \sum_{\lambda\in\Lambda_c\left(\zeta\right)}
    \left[\prod_{\left\{\theta'',\theta'''\right\}\in\lambda}
      -N_{\theta''}N_{\theta'''}
      K_{M_{\theta''},M_{\theta'''}}\left(x_{\theta''}-x_{\theta'''}\right)
    \right]
  \right.
\nonumber\\&&{}
  \left.
    \sum_{\theta''\in\zeta}
    \frac{2N_{\theta''} M_{\theta''}}
    {M_{\theta''}^2+\left(x_{\theta''}-y+si\right)^2}
    \frac{se^{-\beta\left(\frac{2J}{y^2+1}+E\left(\zeta\right)\right)}}
    {s-\left(y-x\right)i}
  \right].
  \label{eq:tmp_f_20}
\end{eqnarray}
On condition that we take a limit $\delta\rightarrow0$ and
$M_{\theta}=1$, $1/(M_{\theta}^2+\left(x_{\theta}-y+si\right)^2)$
yields a pole on the real axis with respect to the variable
$y$.  To detour the pole on the real axis and symmetrize the
integral paths, we use the relation (\ref{eq:synmetrization_integarl})
proved in \ref{sec:synmetrization_integarl}.  Then,
(\ref{eq:tmp_f_20}) becomes
\begin{eqnarray}
&&
  \frac{1}{n!}
  \sum_{\theta\in\Theta\left(n\right)}
  \left[\prod_{\sigma\in\theta}N_{\sigma}!\right]
  \sum_{\zeta\in\bar\Theta(\theta+\left\{n+1\right\})}
  \left[
    \prod_{\theta'\in\zeta}(-1)^{N_{\theta'}-1}
    \frac{\left(N_{\theta'}-1\right)!}{N_{\theta'}}
  \right]
  \frac{N_{\theta_{n+1}}}{\left(N_{\theta_{n+1}}-1\right)!}
\nonumber\\&&{}
  \left[
    \sum_{\zeta'\in\Theta\left(N_{\theta_{n+1}}-1\right)}
    B_{N_{\zeta'}}\prod_{\theta'\in\zeta'}\left(N_{\theta'}-1\right)!
  \right]
  \int_{sym}
  \left[
    \prod_{\theta'\in \zeta}\frac{dx_{\theta'}}{2\pi}
  \right]
  \sum_{\lambda\in\Lambda_c\left(\zeta\right)}
  \frac{se^{-\beta E\left(\zeta\right)}}
  {s-\left(x_{\theta_{n+1}}-x\right)i}
\nonumber\\&&{}
  \left[
    \prod_{\left\{\theta',\theta''\right\}\in\lambda,\quad\theta',\theta''\neq\theta_{n+1}}
    C\left(\theta',\theta''\right))
  \right]
  \left[
    \prod_{\left\{\theta_{n+1},\theta'\right\}\in\lambda}
    E\left(\theta_{n+1},\theta'\right)
  \right].
  \label{eq:tmp_f_03}
\end{eqnarray}
Here, we assume the relation
$\{n+1\}\in\sigma_{n+1}\in\theta_{n+1}\in\zeta$. We also use this
symbol in the same sense as in (\ref{eq:tmp_f_05}),
(\ref{eq:tmp_f_11}) and (\ref{eq:tmp_f_12}).  The constant $B_n$ in
(\ref{eq:tmp_f_03}) is known as the Bernoulli number and is 
determined by
\begin{eqnarray}
  \sum_{n=0}^{\infty}\frac{B_n}{n!}x^n&\equiv&\frac{x}{1-e^{-x}}.
  \label{eq:def_B_n}
\end{eqnarray}
For example, first few $B_n$'s are
\begin{eqnarray}
  B_0=1           \quad,\quad
  B_1=\frac12     \quad,\quad
  B_2=\frac16     \quad,\quad
  B_3=0           \quad,\quad
  B_4=-\frac1{30} \quad,\quad
  \cdots.
\end{eqnarray}
The symbol $\int_{sym}$ in (\ref{eq:tmp_f_03}) indicates that each
path of integration symmetrically avoids poles on the real axis.
Because all the poles on the real axis have the form
$1/(x_\theta-x_{\theta'})$, we can write explicitly $\int_{sym}$ as
\begin{eqnarray}
  \int_{sym}\prod_{n=1}^{N} dx_n
&\equiv&
  \frac1{N!}\sum_{\{p\in P_N\}}\prod_{n=1}^N
  \left[
    \int_{-\infty+n\delta i}^{\infty+n\delta i}
    dx_{p\left(n\right)}
  \right].
  \label{eq:symmetrical_integral}
\end{eqnarray}
The functions $C\left(\theta,\theta'\right)$ and
$E\left(\theta,\theta'\right)$ are defined by
\begin{eqnarray}
  C\left(\theta,\theta'\right)
&\equiv&
  -N_{\theta},N_{\theta'}
  K_{M_{\theta},M_{\theta'}}\left(x_{\theta}-x_{\theta'}\right),
  \label{eq:define_C()}
\\
  E\left(\theta,\theta'\right)
  &\equiv&
  -\left(N_{\theta}-1\right)N_{\theta'}
  K_{M_{\theta},M_{\theta'}}\left(x_{\theta}-x_{\theta'}\right)
  -N_{\theta'}
  K_{M_{\theta}-1,M_{\theta'}}\left(x_{\theta}-x_{\theta'}-si\right)
\nonumber\\&&{}
  -\frac{2N_{\theta'}M_{\theta'}}
  {M_{\theta'}^2+\left(x_{\theta}-x_{\theta'}+s\left(M_{\theta}-2\right)i\right)^2}.
  \label{eq:define_E()}
\end{eqnarray}
Note here that the second term of (\ref{eq:define_E()}) is $0$ for
$E\left(\theta,\theta'\right)$ used in (\ref{eq:tmp_f_03}) because of
the relation $M_{\theta_{n+1}}=1$.  Then, by use of the relation
\begin{eqnarray}
  \sum_{\theta\in\Theta\left(n\right)}
  B_{N_{\theta}}\prod_{\sigma\in\theta}\left(N_\sigma-1\right)!
&=&
  \frac{n!}{n+1},
  \label{eq:combinatorical_relation_2}
\end{eqnarray}
which is easily shown, (\ref{eq:tmp_f_03}) becomes
\begin{eqnarray}
&&
  \frac{1}{n!}
  \sum_{\theta\in\Theta\left(n\right)}
  \left[\prod_{\sigma\in\theta}N_{\sigma}!\right]
  \sum_{\zeta\in\bar\Theta(\theta+\left\{n+1\right\})}
  \left[
    \prod_{\theta'\in\zeta}(-1)^{N_{\theta'}-1}
    \frac{\left(N_{\theta'}-1\right)!}{N_{\theta'}}
  \right]
  \frac{1}{N_{\theta_{n+1}}}
  \int_{sym}
  \left[
    \prod_{\theta'\in \zeta}\frac{dx_{\theta'}}{2\pi}
  \right]
\nonumber\\&&{}
  \sum_{\lambda\in\Lambda_c\left(\zeta\right)}
  \frac{sN_{\theta_{n+1}}e^{-\beta E\left(\zeta\right)}}
  {s-\left(x_{\theta_{n+1}}\!\!-x\right)i}
  \left[
    \prod_{\left\{\theta',\theta''\right\}\in\lambda,\:\theta',\theta''\neq\theta_{n+1}}
    \!\!\!
    C\left(\theta',\theta''\right)
  \right]
  \left[
    \prod_{\left\{\theta_{n+1},\theta'\right\}\in\lambda}
    E\left(\theta_{n+1},\theta'\right)
  \right].
  \label{eq:tmp_f_05}
\end{eqnarray}
Changing the order of the sums, this expression becomes
\begin{eqnarray}
&&
  \frac{1}{\left(n+1\right)!}
  \sum_{\theta\in\Theta\left(n+1\right)}
  \left[\prod_{\sigma\in\theta}N_{\sigma}!\right]
  \sum_{\zeta\in\bar\Theta(\theta)}
  \left[
    \prod_{\theta'\in\zeta}(-1)^{N_{\theta'}-1}
    \frac{\left(N_{\theta'}-1\right)!}{N_{\theta'}}
  \right]
  \int_{sym}
  \left[
    \prod_{\theta'\in \zeta}\frac{dx_{\theta'}}{2\pi}
  \right]
\nonumber\\&&{}
  \sum_{\lambda\in\Lambda_c\left(\zeta\right)}
  \sum_{\theta_0\in\zeta,M_{\theta_0}=1}
  \frac{sN_{\theta_0}e^{-\beta E\left(\zeta\right)}}
  {s-\left(x_{\theta_0}-x\right)i}
  \left[
    \prod_{\left\{\theta',\theta''\right\}\in\lambda,\:\theta',\theta''\neq\theta_0}
    \!\!\!\!\!\!\!\!
    C\left(\theta',\theta''\right))
  \right]\!\!\!
  \left[
  \prod_{\left\{\theta_0,\theta'\right\}\in\lambda}
    \!\!\!
    E\left(\theta_0,\theta'\right)
  \right]\!\!.
\label{eq:tmp_f_04}
\end{eqnarray}
Thus, using the above relations we have shown that (\ref{eq:tmp_f_02})
is expressed as (\ref{eq:tmp_f_04}).

Second, we write the sum of the second and third terms of
(\ref{eq:tmp_f_01}),
\begin{eqnarray}
   -\sum_{s=\pm1}s\sum_{\theta\in\Theta\left(n\right)}\frac{N_\theta!}{n!}
   \int_{C_s}
   \frac{\prod_{\sigma\in\theta}-N_\sigma!u_{N_\sigma}\left(y\right)}
   {2s-\left(y-x\right)i}
   e^{-\frac{2J\beta}{(y+si)^2+1}}
   \frac{dy}{2\pi},
\label{eq:tmp_f_06}
\end{eqnarray}
where, $C_{\pm1}$ means $C_\pm$. In the following change of the
expressions, the sum with respect to $s$ does not play an important
role.  This situation is similar to the one for (\ref{eq:tmp_f_02}).
Then, the symbol $\sum_{s=\pm1}$ will be also omitted for a
while.  Using the relation (\ref{eq:simplification_prod_u_n}), we have
from (\ref{eq:tmp_f_06}),
\begin{eqnarray}
&&
  -\frac{1}{n!}
   \int_{C_s}
    \frac{dy}{2\pi}
  \sum_{\theta\in\Theta\left(n\right)}
  \prod_{\sigma\in\theta}
  \left[
    -\sum_{\theta' \in\Theta(\sigma)}
    \left[\prod_{\sigma'\in\theta'}N_{\sigma'}!\right]
    \sum_{\zeta\in\bar\Theta(\theta')}
    \left[
      \prod_{\theta''\in\zeta}(-1)^{N_{\theta''}-1}
      \frac{\left(N_{\theta''}-1\right)!}
           {N_{\theta''}}
    \right]
  \right.
\nonumber\\&&{}
  \left.
    \left[
      \prod_{\theta''\in \zeta}
      \int_{-\infty}^{\infty}
      \frac{dx_{\theta''}}{2\pi}
      \right]
  \sum_{\lambda\in\Lambda_c\left(\zeta\right)}
  \left[\prod_{\left\{\theta'',\theta'''\right\}\in\lambda}
    -N_{\theta''}N_{\theta'''}
    K_{M_{\theta''},M_{\theta'''}}\left(x_{\theta''}-x_{\theta'''}\right)
  \right]
  \right.
\nonumber\\&&{}
  \left.
     \sum_{\theta''\in\zeta}
     \frac{2N_{\theta''} M_{\theta''}}
          {M_{\theta''}^2+\left(x_{\theta''}-y\right)^2}
      \frac{s e^{-\beta\left(\frac{2J}{\left(y+si\right)^2+1}+E\left(\zeta\right)\right)}}
           {2s-\left(y-x\right)i}
  \right].
\label{eq:tmp_f_21}
\end{eqnarray}
We use a relation,
\begin{eqnarray}
&&
  \int_{C_s}dy
  f\left(y\right)
  \prod_{\sigma\in\theta}
  \left[
    \int_{-\infty}^{\infty}
    \frac{dx_\sigma}{2\pi}
    \frac{2N_\sigma f_{\sigma}\left(x_\sigma\right)}
         {N_\sigma^2+\left(y-x_\sigma\right)^2}
  \right]
\nonumber\\&=&
  \int_{C_s}dy
  f\left(y+msi\right)
  \prod_{\sigma\in\theta}
  \left[
    \int_{-\infty}^{\infty}
    \frac{dx_\sigma}{2\pi}
    \frac{2N_\sigma f_{\sigma}\left(x_\sigma\right)}
         {N_\sigma^2+\left(y-x_\sigma+msi\right)^2}
  \right]
\nonumber\\&&{}
  +\sum_{n=0}^{m-1}
  \int_{C_s}dy
  f\left(y+nsi\right)
  \left[
    \prod_{\sigma\in\theta}
    \int_{-\infty}^{\infty}
    \frac{dx_\sigma}{2\pi}
    \frac{2N_\sigma f_{\sigma}\left(x_\sigma\right)}
         {N_\sigma^2+\left(y-x_\sigma+nsi\right)^2}
\right.
\nonumber\\&&
\left.{}
    -\prod_{\sigma\in\theta}
    \left[
      \int_{-\infty}^{\infty}
      \frac{dx_\sigma}{2\pi}
      \frac{2N_\sigma f_{\sigma}\left(x_\sigma\right)}
           {N_\sigma^2+\left(y-x_\sigma+nsi\right)^2}
      -
      \delta\left(n+1,N_\sigma\right)
      f_{\sigma}\left(y-si\right)
    \right]
  \right],
\label{eq:tmp_f_09}
\end{eqnarray}
where $f_\sigma(x)$ and $f(x)$ are analytic functions in the regions
$-1<\Im x\leq 0$ and $0<\Im x \leq m$ in case of $s=1$ ($0\leq\Im x<1$ and $-m\leq\Im x<0$ in case of $s=-1$) respectively, and
$\max_{\sigma\in\theta}N_\sigma=m$.  Then, (\ref{eq:tmp_f_21}) becomes
\begin{eqnarray}
&&
  \frac{1}{n!}
  \sum_{\theta\in\Theta\left(n\right)}
  \left[\prod_{\sigma\in\theta}N_{\sigma}!\right]
  \sum_{\zeta\in\bar\Theta(\theta)}
  \left[
    \prod_{\theta'\in\zeta}(-1)^{N_{\theta'}-1}
    \frac{\left(N_{\theta'}-1\right)!}{N_\theta'}
  \right]
  \sum_{\theta'\in\zeta}
  \frac{N_{\theta'}\left(-1\right)^{N_{\theta'}-1}}{\left(N_{\theta'}-1\right)!}
\nonumber\\&&{}
  \left[
    \sum_{\theta''\in\Theta\left(N_{\theta'}\right)}
    \prod_{\sigma''\in\theta''}
    \left(-1\right)^{N_{\sigma''}-1}
    \left(N_{\sigma''}-1\right)!
  \right]
   \int_{C_s}
    \frac{dx_{\theta'}}{2\pi}
    \left[
      \prod_{\theta''\in \zeta, \theta''\neq\theta'}
      \int_{-\infty}^{\infty}
      \frac{dx_{\theta''}}{2\pi}
    \right]
\nonumber\\&&{}
  \sum_{\lambda\in\Lambda_c\left(\zeta\right)}
  \!\!
  \frac{
        s \exp\left[-\beta\left(
            E\left(\zeta\right)
            +\frac{2J}{\left(x_{\theta'}+sM_{\theta'}i\right)^2+1}
            +\frac{2JM_{\theta'}}{\left(x_{\theta'}-si\right)^2+M_{\theta'}^2}
            -\frac{2JM_{\theta'}}{x_{\theta'}^2+M_{\theta'}^2}
          \right)\right]}
           {\left(M_{\theta'}+1\right)s-\left(x_{\theta'}-x\right)i}
           \!
  \left[
    \prod_{\left\{\theta'',\theta'''\right\}\in\lambda,
      \theta'',\theta'''\neq\theta'}
    \!\!\!\!\!\!\!\!\!\!\!
    C\left(\theta'',\theta'''\right)
    \!\!
  \right]
\nonumber\\&&{}
\!\!
  \prod_{\left\{\theta',\theta''\right\}\in\lambda}
  \left[
    -N_{\theta'}N_{\theta''}
    K_{M_{\theta'},M_{\theta''}}\left(x_{\theta'}-x_{\theta''}-si\right)
    -\frac{2N_{\theta''} M_{\theta''} }{M_{\theta''}^2
      +\left(x_{\theta'}-x_{\theta''}+s\left(M_{\theta'}-1\right)i\right)^2}
  \right]\!\!.
\makebox[1cm]{}
\label{eq:tmp_f_22}
\end{eqnarray}
We point out that in this case the first term of the r.h.s. in
(\ref{eq:tmp_f_09}) is $0$ because there is no pole inside (outside)
the path of integration $C_+$ ($C_-$) with respect to $y$ including
the point $|y|=\infty$.  Then, using a simple relation $
\sum_{\theta\in\Theta_n} \prod_{\sigma\in\theta}
\left(-\right)^{N_\sigma-1} \left(N_\sigma-1\right)!  =\delta_{n,1}$,
we get from (\ref{eq:tmp_f_22}),
\begin{eqnarray}
&&
  \frac{1}{n!}
  \sum_{\theta\in\Theta\left(n\right)}\!
  \left[\prod_{\sigma\in\theta}N_{\sigma}!\right]\!
  \sum_{\sigma_0\in\theta}
  \sum_{\zeta\in\bar\Theta(\theta-\sigma_0)}
  \!\!\!
  \left[
    \prod_{\theta'\in\zeta}(-1)^{N_{\theta'}-1}
    \frac{\left(N_{\theta'}-1\right)!}{N_{\theta'}}
  \right]
  \!
  \int_{-\infty+s\delta i}^{\infty+s\delta i}
  \frac{dx_{\theta_0}}{2\pi}
  \left[
    \prod_{\theta'\in \zeta}
    \int_{-\infty}^{\infty}
    \frac{dx_{\theta'}}{2\pi}
  \right]
\nonumber\\&&{}
  \!\!\!
  \!\!\!
  \sum_{\lambda\in\Lambda_c\left(\zeta+\theta_0\right)}
  \frac{
    s \exp\left[-\beta\left(
        E\left(\zeta\right)
        +\frac{2J\left(M_{\theta_0}+1\right)}
        {x_{\theta_0}^2+\left(M_{\theta_0}+1\right)^2}
        -\frac{2JM_{\theta_0}}{x_{\theta_0}^2+M_{\theta_0}^2}
      \right)\right]}
       {\left(M_{\theta_0}+1\right)s-\left(x_{\theta_0}-x\right)i}
  \left[
    \prod_{\left\{\theta',\theta''\right\}\in\lambda,
      \theta',\theta''\neq\theta_0}
    C\left(\theta',\theta''\right)
  \right]
\nonumber\\&&{}
  \prod_{\left\{\theta_0,\theta'\right\}\in\lambda}
  \left[
    -N_{\theta'}
    K_{M_{\theta_0},M_{\theta'}}\left(x_{\theta_0}-x_{\theta'}-si\right)
    -\frac{2N_{\theta'} M_{\theta'} }{M_{\theta'}^2
      +\left(x_{\theta_0}-x_{\theta'}+s\left(M_{\theta_0}-1\right)i\right)^2}
  \right],
\label{eq:tmp_f_13}
\end{eqnarray}
where $\theta_0\equiv\{\sigma_0\}$.
Here, we have changed the path of integration from $C_s$ to
${-\infty+s\delta i}\rightarrow{\infty+s\delta i}$, because there is
no pole at $|x_{\theta_0}|\rightarrow\infty$.  In
eq.(\ref{eq:tmp_f_13}), when we take a limit $\delta\rightarrow0$ and
$M_{\theta}=M_{\theta_0}+1$, the last term contains a pole on the real
axis with respect to $x_{\theta_0}$.  Symmetrizing by use of
(\ref{eq:synmetrization_integarl}) the integral paths which avoid the
pole on the real axis, we obtain from (\ref{eq:tmp_f_13}),
\begin{eqnarray}
&&
  \frac{1}{n!}
  \sum_{\theta\in\Theta\left(n+1\right),N_{\sigma_{n+1}}\neq1}
  \left[\prod_{\sigma\in\theta}N_{\sigma}!\right]
  \sum_{\zeta\in\bar\Theta(\theta)}
  \left[
    \prod_{\theta'\in\zeta}(-1)^{N_{\theta'}-1}
    \frac{\left(N_{\theta'}-1\right)!}{N_{\theta'}}
  \right]
  \frac{N_{\theta_{n+1}}}{N_{\sigma_{n+1}}\left(N_{\theta_{n+1}}-1\right)!}
\nonumber\\&&{}
  \left[
    \sum_{\theta'\in\Theta\left(N_{\theta_{n+1}}-1\right)}
    B_{N_{\theta'}}
    \prod_{\sigma'\in\theta'}
    {\left(N_{\sigma'}-1\right)!}
  \right]
    \int_{sym}
    \left[
      \prod_{\theta'\in \zeta}
      \frac{dx_{\theta'}}{2\pi}
    \right]
\sum_{\lambda\in\Lambda_c\left(\zeta\right)}
      \frac{
        s e^{ E\left(\zeta\right)}
          }
           {M_{\theta_{n+1}}s-\left(x_{\theta_{n+1}}-x\right)i}
\nonumber\\&&{}
  \left[
    \prod_{\left\{\theta',\theta''\right\}\in\lambda,
      \quad\theta',\theta''\neq\theta_{n+1}}
    C\left(\theta',\theta''\right)
  \right]
  \left[
  \prod_{\left\{\theta_{n+1},\theta'\right\}\in\lambda}
    E\left(\theta_{n+1},\theta'\right)
  \right]
\label{eq:tmp_f_11}.
\end{eqnarray}
By use of  (\ref{eq:combinatorical_relation_2}), (\ref{eq:tmp_f_11}) is
written as
\begin{eqnarray}
&&
  \frac{1}{n!}
  \sum_{\theta\in\Theta\left(n+1\right),N_{\sigma_{n+1}}\neq1}
  \!\!
  \left[\prod_{\sigma\in\theta}N_{\sigma}!\right]
  \!\!
  \sum_{\zeta\in\bar\Theta(\theta)}
  \!\!
  \left[
    \prod_{\theta'\in\zeta}(-1)^{N_{\theta'}-1}
    \frac{\left(N_{\theta'}-1\right)!}{N_{\theta'}}
  \right]
  \frac{1}{N_{\sigma_{n+1}}N_{\theta_{n+1}}}
    \int_{sym}
    \left[
      \prod_{\theta'\in \zeta}
      \frac{dx_{\theta'}}{2\pi}
    \right]
\nonumber\\&&{}
\sum_{\lambda\in\Lambda_c\left(\zeta\right)}
      \frac{
        sN_{\theta_{n+1}} e^{ E\left(\zeta\right)}
          }
           {M_{\theta_{n+1}}s-\left(x_{\theta_{n+1}}-x\right)i}
  \left[
    \prod_{\left\{\theta',\theta''\right\}\in\lambda,
      \theta',\theta''\neq\theta_{n+1}}
\!\!\!\!    C\left(\theta',\theta''\right)
  \right]
  \left[
    \prod_{\left\{\theta_{n+1},\theta'\right\}\in\lambda}
\!\!\!    E\left(\theta_{n+1},\theta'\right)
  \right].
\label{eq:tmp_f_12}
\end{eqnarray}
Changing the order of the summations, this expression becomes
\begin{eqnarray}
&&
  \frac{1}{\left(n+1\right)!}
  \sum_{\theta\in\Theta\left(n+1\right)}
  \left[\prod_{\sigma\in\theta}N_{\sigma}!\right]
  \sum_{\zeta\in\bar\Theta(\theta)}
  \left[
    \prod_{\theta'\in\zeta}(-1)^{N_{\theta'}-1}
    \frac{\left(N_{\theta'}-1\right)!}{N_{\theta'}}
  \right]
    \int_{sym}
    \left[
      \prod_{\theta'\in \zeta}
      \frac{dx_{\theta'}}{2\pi}
    \right]
\nonumber\\&&{}
\makebox[-2mm]{}
  \sum_{\lambda\in\Lambda_c\left(\zeta\right)}
  \sum_{\theta_0\in\zeta,M_{\theta_0}\neq1}
  \!\!\!\!
      \frac{
        sN_{\theta_0} e^{ E\left(\zeta\right)}
          }
           {M_{\theta_0}s-\left(x_{\theta_0}-x\right)i}
  \left[
    \prod_{\left\{\theta',\theta''\right\}\in\lambda,
      \:\theta',\theta''\neq\theta_0}
    \!\!\!\!\!\!\!\!
    C\left(\theta',\theta''\right)
  \right]\!\!\!\!
  \left[
  \prod_{\left\{\theta_0,\theta'\right\}\in\lambda}
    \!\!\!
    E\left(\theta_0,\theta'\right)
  \right]\!\!.
\label{eq:tmp_f_07}
\end{eqnarray}
From the above relations we see that (\ref{eq:tmp_f_06}) is
equal to (\ref{eq:tmp_f_07}).

Both parts of the r.h.s. of (\ref{eq:tmp_f_01}) are replaced with
(\ref{eq:tmp_f_05}) and (\ref{eq:tmp_f_07}). Then, the r.h.s. of
(\ref{eq:tmp_f_01}) is expressed as
\begin{eqnarray}
&&  \frac{1}{\left(n+1\right)!}
  \sum_{\theta\in\Theta\left(n+1\right)}
  \left[\prod_{\sigma\in\theta}N_{\sigma}!\right]
  \sum_{\zeta\in\bar\Theta(\theta)}
  \left[
    \prod_{\theta'\in\zeta}(-1)^{N_{\theta'}-1}
    \frac{\left(N_{\theta'}-1\right)!}{N_{\theta'}}
  \right]
    \int_{sym}
    \left[
      \prod_{\theta'\in \zeta}
      \frac{dx_{\theta'}}{2\pi}
    \right]
\nonumber\\&&{}
\sum_{\lambda\in\Lambda_c\left(\zeta\right)}
\sum_{\theta_0\in\zeta}
\sum_{s=\pm1}
      \frac{
        sN_{\theta_0} e^{ E\left(\zeta\right)}
          }
           {M_{\theta_0}s-\left(x_{\theta_0}-x\right)i}
  \left[
    \prod_{\left\{\theta',\theta''\right\}\in\lambda,
      \:\theta',\theta''\neq\theta_0}
    \!\!\!\!\!\!\!\!
    C\left(\theta',\theta''\right)
  \right]
  \!\!\!\!
  \left[
  \prod_{\left\{\theta_0,\theta'\right\}\in\lambda}
    \!\!\!
    E\left(\theta_0,\theta'\right)
  \right]\!.
\label{eq:tmp_f_08}
\end{eqnarray}
Using eq.(\ref{eq:ano_rep_modified_jacobian}) proved in
\ref{sec:ano_rep_modified_jacobian}, we can confirm the definition of $u_{n+1}(x)$, which is the
r.h.s. of (\ref{eq:relation_u_n}).

\section{
An expression of $u_n(x)$
}
\label{sec:proof_wrt_u_n}
In this appendix, we prove the following expression of  $u_n(x)$,
\begin{eqnarray}
&&
  \sum_{\theta\in\Theta\left(n\right)}\frac{N_\theta!}{n!}
  \prod_{\sigma\in\theta}-N_\sigma!u_{N_\sigma}\left(y\right)
\nonumber\\
&=&
  \frac{1}{n!}
  \sum_{\theta\in\Theta\left(n\right)}
  \prod_{\sigma\in\theta}
  \left[
    -\sum_{\theta' \in\Theta(\sigma)}
    \left[\prod_{\sigma'\in\theta'}N_{\sigma'}!\right]
  \right.
\nonumber\\&&{}
  \left.
    \sum_{\zeta\in\bar\Theta(\theta')}
    \left[
      \prod_{\theta''\in\zeta}(-1)^{N_{\theta''}-1}\left(N_{\theta''}-1\right)!
    \right]
    \int
    \left|\frac{\partial I}{\partial x}\right|_{c1,\zeta}\left(y\right)
    e^{-\beta E\left(\zeta\right)} \prod_{\theta'\in \zeta}d
    x_{\theta'}
  \right].
\label{eq:simplification_prod_u_n}
\end{eqnarray}
Here, the definition of $u_n\left(x\right)$ is
(\ref{eq:definition_un}), and the definition of
$\left|\frac{\partial I}{\partial x}\right|_{c1,\zeta}(y)$ is
\begin{eqnarray}
&&    \left({2\pi}
    \right)^{N_{\zeta}}
  \left|\frac{\partial I}{\partial x}\right|_{c1,\zeta}\left(y\right)
\nonumber\\
&\equiv&
  \left[
    \prod_{\theta\in\zeta}N_\theta\right]^{-1}
  \sum_{\lambda\in\Lambda_c\left(\zeta\right)}
  \left[\prod_{\left\{\theta,\theta'\right\}\in\lambda}
    -N_{\theta}N_{\theta'}
    K_{M_{\theta},M_{\theta'}}\left(x_{\theta}-x_{\theta'}\right)
  \right]
     \sum_{\theta\in\zeta}
     \frac{2N_\theta M_\theta}{M_\theta^2+\left(x_\theta-y\right)^2}.
\end{eqnarray}
Substituting (\ref{eq:definition_un}) for
$u_{N\sigma}(y)$ in the l.h.s.  of (\ref{eq:simplification_prod_u_n})
and using a relation
\begin{eqnarray}
  \left|\frac{\partial I}{\partial x}\right|_{1,\zeta}\left(y\right)
&=&
  \sum_{\xi\in\Theta\left(\zeta\right)}\prod_{\zeta'\in\xi}
  \left|\frac{\partial I}{\partial x}\right|_{c1,\zeta'}\left(y\right),
\end{eqnarray}
which is proved in the same way as (\ref{eq:struct_jacobian}), we can
show that the l.h.s. of (\ref{eq:simplification_prod_u_n}) is
calculated as
\begin{eqnarray}
&&  \sum_{\theta\in\Theta\left(n\right)}\frac{N_\theta!}{n!}
  \prod_{\sigma\in\theta}
  \left[
    -\sum_{\theta' \in\Theta(\sigma)}
    \left[\prod_{\sigma'\in\theta'}N_{\sigma'} !\right]
    \sum_{\zeta\in\bar\Theta(\theta')}
    \left[
      \prod_{\theta''\in\zeta}(-1)^{N_{\theta''}-1}\left(N_{\theta''}-1\right)!
    \right]
  \right.
\nonumber\\&&{}
  \left.
  \sum_{\xi\in\Theta\left(\zeta\right)}
  \left[
    \prod_{\zeta'\in\xi}
    \int
    \left|\frac{\partial I}{\partial x}\right|_{c1,\zeta'}\left(y\right)
    e^{-\beta E\left(\zeta'\right)} \prod_{\theta'\in \zeta}d
    x_{\theta'}
  \right]
  \right]
\nonumber\\&=&
  \frac{1}{n!}
  \sum_{\theta\in\Theta\left(n\right)}
  \left[
    \sum_{\theta'\in\Theta\left(N_\theta\right)}
    N_{\theta'}!\left(-1\right)^{N_{\theta'}}
  \right]
  \prod_{\sigma\in\theta}
  \left[
    \sum_{\theta' \in\Theta(\sigma)}
    \left[\prod_{\sigma'\in\theta'}N_{\sigma'}!\right]
  \right.
\nonumber\\&&{}
  \left.
    \sum_{\zeta\in\bar\Theta(\theta')}
    \left[
      \prod_{\theta''\in\zeta}(-1)^{N_{\theta''}-1}\left(N_{\theta''}-1\right)!
    \right]
    \int
    \left|\frac{\partial I}{\partial x}\right|_{c1,\zeta}\left(y\right)
    e^{-\beta E\left(\zeta\right)} \prod_{\theta'\in \zeta}d
    x_{\theta'}
  \right].
\label{eq:tmp_i_01}
\end{eqnarray}
In this equality, we have changed the order of the sums.  Using a
relation
\begin{eqnarray}
  \sum_{\theta\in\Theta\left(n\right)}
  N_\theta!\left(-1\right)^{N_\theta}
&=&
  \left(-1\right)^{n}
\end{eqnarray}
for simplification of the first term of (\ref{eq:tmp_i_01}), we obtain
the r.h.s. of (\ref{eq:simplification_prod_u_n}).

\section{
Symmetrization of the paths of integrals
}
\label{sec:synmetrization_integarl}
In this appendix, we prove a relation,
\begin{eqnarray}
&&
  \int_{-\infty+\delta i}^{\infty+\delta i}
   dx f_0\left(x\right)
  \prod_{n=1}^{N}
  \left[
    \int_{-\infty}^{\infty}
    \frac{dy_n}{2\pi i}
    \frac{1}{y_n-x}f_n\left(y_n\right)
  \right]
\nonumber\\&=&
  \sum_{\sigma\subseteq\{1,\cdots,N\}}B_{N_\sigma}\int_{sym}
  f_0\left(x\right)
  \left[
    \prod_{n\in\sigma}f_n\left(x\right)
  \right]
  \left[
    \prod_{n\in\{1,\cdots,N\}-\sigma}
    \frac{1}{y_n-x}f_n\left(y_n\right)
    \frac{dy_n}{2\pi i}
  \right]
  dx.
\label{eq:synmetrization_integarl}
\end{eqnarray}
Here, the definitions of $B_n$ and $\int_{sym}$ are in
(\ref{eq:def_B_n}) and (\ref{eq:symmetrical_integral}) respectively,
and we assume that $f_n(x)$ is an analytic function on the real axis.

The paths of integrations in the r.h.s. of
(\ref{eq:synmetrization_integarl}) are changed into ones in the l.h.s.
of (\ref{eq:synmetrization_integarl}). In other words, symmetrical
paths of integrations are changed into ones which keep the relation
$\Im y_n<\Im x$.  Then, the r.h.s. of
(\ref{eq:synmetrization_integarl}) becomes
\begin{eqnarray}
&&
\sum_{\sigma\subseteq\{1,\cdots,N\}}
  \left[
    \sum_{m=0}^{N_\sigma}
    \left(-1\right)^{N_\sigma-m}
      \frac{N_\sigma!}{\left(N_\sigma-m\right)!m!}
    \left[
      \sum_{l=0}^{N-N_\sigma}
      \frac{\left(l+N_\sigma-m\right)!}{l!}
      \frac{\left(N-N_\sigma\right)!}{\left(N-m+1\right)!}
    \right]
    B_m
  \right]
\nonumber\\&&{}
  \int_{-\infty+\delta i}^{\infty+\delta i} 
  dx
  f_0\left(x\right)
  \left[
    \prod_{n\in\sigma}f_n\left(x\right)
  \right]
    \prod_{n\in\{1,\cdots,N\}-\sigma}
  \left[
    \int_{-\infty}^{\infty}
    \frac{dy_n}{2\pi i}
    \frac{1}{y_n-x}f_n\left(y_n\right)
  \right].
\label{eq:tmp_g_01}
\end{eqnarray}
Using the following two relations which are easily shown,
\begin{eqnarray}
      \sum_{l=0}^{n}
      \frac{\left(l+m\right)!}{l!}
      &=&\frac{\left(n+m+1\right)!}{n!\left(m+1\right)}
\end{eqnarray}
\begin{eqnarray}
  \sum_{m=0}^{n}\left(-1\right)^{n-m}\frac{n!}{\left(n-m+1\right)!m!}B_m
  &=&
  \delta_{n,0}
\end{eqnarray}
for (\ref{eq:tmp_g_01}), we obtain the the l.h.s. of
(\ref{eq:synmetrization_integarl}).

\section{ 
Alternative representation of the modified Jacobian
}
\label{sec:ano_rep_modified_jacobian}
In this appendix, we show that
\begin{eqnarray}
\label{eq:ano_rep_modified_jacobian}
    \left(
      {2\pi}
    \right)^{N_{\zeta}}
  \left|\frac{\partial I}{\partial x}\right|_{1,\zeta}
\left(y\right)
&=&
  \left[\prod_{\theta\in\zeta}N_\theta\right]^{-1}
  \sum_{\lambda\in\Lambda_c\left(\zeta\right)}
  \sum_{\theta\in\zeta}
  \sum_{s=1,-1}
  \frac{sN_\theta}{sM_\theta-\left(x_\theta-y\right)i}
\nonumber\\&&{}
  \left[
    \prod_{\left\{\theta',\theta''\right\}\in\lambda \quad \theta',\theta''\neq\theta}
    C\left(\theta',\theta''\right)
  \right]
  \left[
  \prod_{\left\{\theta,\theta'\right\}\in\lambda}
  E\left(\theta,\theta'\right)
  \right]
\end{eqnarray}
is an alternative representation of the modified Jacobian
(\ref{eq:def_modified_jacobian}).  Here,
$C\left(\theta,\theta'\right)$ and $E\left(\theta,\theta'\right)$ are
defined in (\ref{eq:define_C()}) and (\ref{eq:define_E()})
respectively.

We can rewrite (\ref{eq:def_modified_jacobian}) as
\begin{eqnarray}
&&    \left(
      {2\pi}
    \right)^{N_{\zeta}}
  \left|\frac{\partial I}{\partial x}\right|_{1,\zeta}\left(y\right)
\nonumber\\&=&
  \left[\prod_{\theta\in\zeta}N_\theta\right]^{-1}
  \sum_{\lambda\in\Lambda_c\left(\zeta\right)}
  \sum_{\theta\in\zeta}
  \left[
    \prod_{\left\{\theta',\theta''\right\}\in\lambda \quad \theta',\theta''\neq\theta}
    C\left(\theta',\theta''\right)
  \right]
     \frac{2N_\theta M_\theta}{M_\theta^2+\left(x_\theta-y\right)^2}
\nonumber\\&&{}
  \sum_{\zeta'\subseteq\{\theta'|\{\theta,\theta'\}\in\lambda\}}
  \frac1{N_{\zeta'}+1}
  \left[
    \prod_{\theta'\in\{\theta'|\{\theta,\theta'\}\in\lambda\}-\zeta'}
    C\left(\theta,\theta'\right)
  \right]
  \left[
    \prod_{\theta'\in\zeta'}
     \frac{2N_{\theta'} M_{\theta'}}{M_{\theta'}^2+\left(x_{\theta'}-y\right)^2}
  \right]
\nonumber\\&=&
  \left[\prod_{\theta\in\zeta}N_\theta\right]^{-1}
  \sum_{\lambda\in\Lambda_c\left(\zeta\right)}
  \sum_{\theta\in\zeta}
  \sum_{s=1,-1}
  \frac{sN_\theta}{sM_\theta-\left(x_\theta-y\right)i}
  \left[
    \prod_{\left\{\theta',\theta''\right\}\in\lambda \quad \theta',\theta''\neq\theta}
    C\left(\theta',\theta''\right)
  \right]
\nonumber\\&&{}
  \prod_{\{\theta,\theta'\}\in\lambda}
  \left[
    C\left(\theta,\theta'\right)
    +
  \frac{2N_{\theta'}M_{\theta'}}
    {M_{\theta'}^2+\left(x_{\theta'}-x_\theta-s M_\theta\right)^2}
  \right].
\label{eq:tmp_d_01}
\end{eqnarray}
In the first equality, we have changed the order of the sums.  In the
second equality, we have used the relation
(\ref{eq:functional_relation}) proved in
\ref{sec:complex_fractional_relation}. Applying a relation
\begin{eqnarray}
&&
    K_{M_{\theta},M_{\theta'}}\left(x_{\theta}-x_{\theta'}\right)
    -\frac{2M_{\theta'}}{M_{\theta'}^2+\left(x_{\theta'}-x_{\theta}-sM_{\theta}i\right)^2}
\nonumber\\&=&
    K_{M_{\theta}-1,M_{\theta'}}\left(x_{\theta}-x_{\theta'}-si\right)
    +\frac{2M_{\theta'}}{M_{\theta'}^2+\left(x_{\theta}-x_{\theta'}+s\left(M_{\theta}-2\right)i\right)^2}
\label{eq:tmp_d_02}
\end{eqnarray}
to (\ref{eq:tmp_d_01}), we obtain the r.h.s. of
(\ref{eq:ano_rep_modified_jacobian}).  In the relation
(\ref{eq:tmp_d_02}), it is supposed that $s$ is $\pm1$.

\section{
 A Proof of a fractional relation
}
\label{sec:complex_fractional_relation}
In this appendix, we prove
\begin{eqnarray}
&&
  \sum_{n=1}^N
  \sum_{\sigma\subseteq\{1,\cdots,N\}-\{n\}}
  \frac1{N-N_\sigma}
  \left[
    \prod_{m\in\sigma}
    c_m
  \right]
  \left[
    \prod_{m\subseteq\{1,\cdots,N\}-\sigma}
    \frac{2a_m d_m}{a_m^2 + (y-x_m)^2}
  \right]
\nonumber\\&=&
  \sum_{s=\pm1}
  \sum_{n=1}^{N}
  \frac{s d_n}{sa_n - \left(x_n-y\right)i}
  \prod_{m\neq n}
  \left[c_m+
    \frac{2a_md_m}{a_m^2+\left(x_m-x_n-s a_ni\right)^2}
  \right],
\label{eq:functional_relation}
\end{eqnarray}
where $a_n$, $c_n$, $d_n$, $x_n$ and $y$ are arbitrary numbers.

 We first prove a relation
\begin{eqnarray}
  1
&=&
  \sum_{n=1}^{N}
  \prod_{m\neq n}\frac{s_ma_m-\left(x_m-y\right)i}{\left(s_ma_m-s_na_n\right)-\left(x_m-x_n\right)i}.
\end{eqnarray}
We regard the r.h.s. as a polynomial of degree $N-1$
with respect to $y$. It is clear that this relation holds on the $N$
points $y=x_n+s_na_ni$. Therefore, this equation is nothing but an
identical equation.  Dividing both sides of this identical equation by
$\prod_{n}[a_n-s_n\left(x_n-y\right)i]d_n^{-1}$, we get
\begin{eqnarray}
  \prod_{n}\frac{s_nd_n}{s_na_n-\left(x_n-y\right)i}
&=&
  \sum_{n=1}^{N}\frac{s_nd_n}{s_na_n-\left(x_n-y\right)i}
  \prod_{m\neq n}\frac{s_md_m}{\left(s_ma_m-s_na_n\right)-\left(x_m-x_n\right)i}.
\end{eqnarray}
Here, we restrict that $s_n$ is $\pm1$.  The sum of each side of this
equation with respect to $\{s_n\}$ gives
\begin{eqnarray}
  \prod_{n}\frac{2a_nd_n}{a_n^2 + (y-x_n)^2}
&=&
  \sum_{s=\pm1}
  \sum_{n=1}^{N}
  \frac{sd_n}{sa_n - \left(x_n-y\right)i}
  \prod_{m\neq n}
  \frac{2a_md_m}{a_m^2+\left(x_m-x_n-sa_ni\right)^2}.
\label{eq:tmp_e_01}
\end{eqnarray}
We are in a position to prove the main issue of this Appendix. Using (\ref{eq:tmp_e_01}) for the l.h.s. of
(\ref{eq:functional_relation}), we get
\begin{eqnarray}
  \sum_{s=\pm1}
\!
  \sum_{n=1}^N
  \frac{sd_n}{sa_n - \left(x_n-y\right)i}
    \!\!\!\!
  \sum_{\sigma\subseteq\{1,\cdots,N\}-\{n\}}
  \left[
    \prod_{m\in\sigma}
    c_m
  \right]
    \!\!\!
  \left[
    \prod_{m \subseteq\{1,\cdots,N\}-\{n\}-\sigma}
    \!\!
    \frac{2a_md_m}{a_m^2+\left(x_m-x_n-sa_ni\right)^2}
  \right]\!\!\!.
\label{eq:tmp_e_02}
\end{eqnarray}
  Expanding the r.h.s. of
(\ref{eq:functional_relation}), we also get (\ref{eq:tmp_e_02}).  This
completes a proof of (\ref{eq:functional_relation}).

\end{document}